\documentclass[prb, twocolumn, showpacs]{revtex4-1}
\usepackage{float}
\usepackage{bbm}
\usepackage{epsfig}
\usepackage{epstopdf}
\usepackage{graphicx, subfigure}
\usepackage{amsmath,amssymb}
\usepackage{color}
\usepackage{hyperref}
\begin{document}

\title{Directly accessible entangling gates for capacitively coupled singlet-triplet qubits}

\author{F.~A.~Calderon-Vargas}
\email{f.calderon@umbc.edu}
\author{J.~P.~Kestner}
\affiliation{Department of Physics, University of Maryland Baltimore County, Baltimore, MD 21250, USA}

\begin{abstract}
The recent experimental advances in capacitively coupled singlet-triplet qubits, particularly the demonstration of entanglement, opens the question of what type of entangling gates the system's Hamiltonian can produce directly via a single square pulse. We address this question by considering the system's Hamiltonian from first principles and using the representation of its nonlocal properties in terms of local invariants. In the analysis we include the three different ways in which the system can be biased and their effect on the generation of entangling gates. We find that, in one of the possible biasing modes, the Hamiltonian has an especially simple form, which can directly generate a wide range of different entangling gates including the iSWAP gate. Moreover, using the complete form of the Hamiltonian we find that, for any biasing mode, a CNOT gate can be generated directly.
\end{abstract}

\pacs{03.67.Lx, 73.21.La, 85.35.Be, 03.67.Bg}

\maketitle

\section{Introduction}
Among the different candidates for a quantum computer, spin qubits in semiconductor quantum dots present the advantages of scalability, long coherence times, and rapid gate operations. In particular, a qubit encoded in the low-lying singlet-triplet subspace of two electrons in a double dot has the promising features of immunity to homogeneous fluctuations of the magnetic field and purely electrical controllability\cite{Levy2002,Petta2005,Foletti2009,Bluhm2010a}. The realization of any quantum logic circuit requires single-qubit and two-qubit operations. In the case of singlet-triplet qubits, two-qubit gates may be implemented by controllable exchange coupling between two
dots from different singlet-triplet qubits \cite{Levy2002,Li2012,Klinovaja2012},or by capacitive coupling with no interqubit tunneling \cite{Taylor2005,Stepanenko2007,Ramon2010,Ramon2011}. The latter
was successfully implemented  in the laboratory \cite{VanWeperen2011,Shulman2012}, including the demonstration of entanglement by using a Hahn echo-like sequence to produce a CPHASE gate \cite{Shulman2012}.\\
\indent The generation of an entangling gate by means of a sequence of two-qubit operations generally complicates the characteristics of the gate error. While errors can often be corrected by pulse sequence techniques \cite{Khodjasteh2009a,Khodjasteh2012,Kestner2013,Wang2014}, it is useful in that context to explicitly consider the properties of a single element of such a sequence. Of particular interest is the question of whether and how maximally entangling operations can be generated directly by evolution under a single constant segment of a specific form of Hamiltonian, information that is enclosed in its \textit{nonlocal} content \cite{Kraus2001,Khaneja2001a,Makhlin2002,Zhang2003}. The characterization of the nonlocal properties of any two-qubit operation is represented by a set of local invariants \cite{Makhlin2002,Zhang2003}, thus a pair of two-qubit operations that have equal values of those invariants are equivalent up to local (single-qubit) operations.\\
\indent In this work we consider, from first principles, the Hamiltonian of two capacitively coupled singlet-triplet qubits  and investigate its entangling properties, specifically its capability to directly generate a maximally entangling gate. In doing so, we consider the different ways in which the four-electron system can be biased and the respective effects on the type of entangling gates that can be generated directly. The paper is divided into four sections. In Sec. \ref{sec:Two-qubit Hamiltonian}, we introduce the two-qubit Hamiltonian, leaving the details of its derivation to the appendix section. Then, in Sec. \ref{sec:Local_invariants}, we present some basics about the characterization of the nonlocal properties of two-qubit operations. Finally, in Sec. \ref{sec:Charge_configuration_entangling_gates} we present the different biasing modes employed in the analysis and the type of entangling gates that can be effectively generated with a single implementation of the two-qubit Hamiltonian.\\

\section{Two-qubit Hamiltonian}\label{sec:Two-qubit Hamiltonian}
\begin{figure}[tbp]
    \centering
    \includegraphics[width=8.5cm]{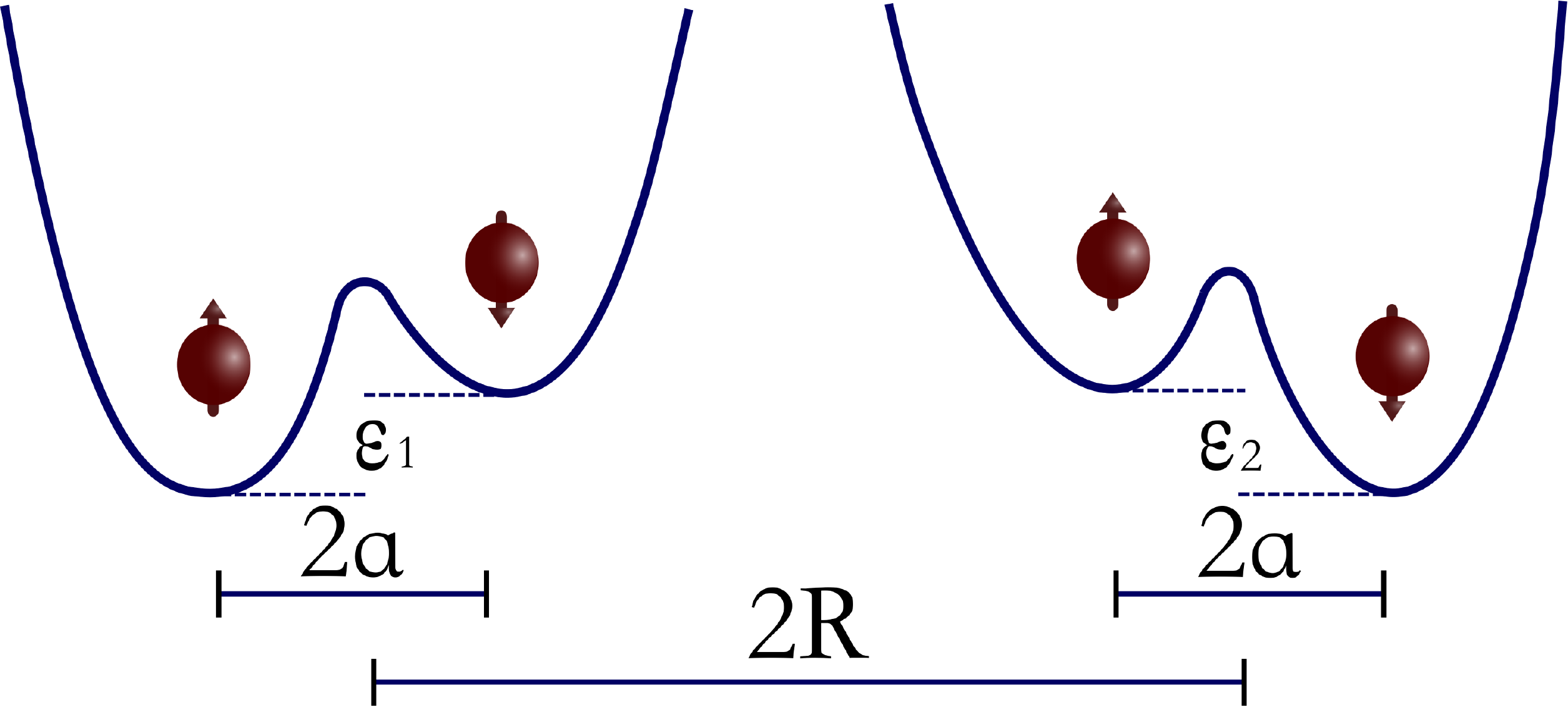}
    \caption{Schematic of two adjacent double quantum dots (DQD) separated by a distance $2R$ and with an interdot distance equal to $2a$. The confinement potential of each DQD is modeled as a quartic function  centered at $(\pm R,0)$. The energy difference between single-dot ground orbitals in the $i$-th DQD is represented by $\varepsilon_i$.}\label{fig:1}
\end{figure}

The system under consideration consists of two adjacent double quantum dots (DQD) separated by a distance $2R$ and with an interdot distance equal to $2a$ (see Fig. \ref{fig:1}). Each DQD confines two electrons in the lowest orbital level of each dot (an external magnetic field makes excited orbital states energetically inaccessible) where the qubit is encoded by the singlet and triplet states of the two electrons. Tunneling between dots within each qubit is controlled by adjusting the bias $\varepsilon$, which is proportional to the voltage difference between the right and left dot electrical gates. Interqubit tunneling is not allowed and the coupling is purely capacitive. In the following, we take the standard approach used in Refs. \onlinecite{Burkard1999,Stepanenko2007,Ramon2011}, except that we consider an external magnetic field in the plane of the two-dimensional electron gas (2DEG), as appropriate for the system we are considering \cite{Shulman2012}, and consequently we do not include the magnetic-based phase factor in the harmonic ground-state wave functions (see Appendix \ref{app:Single-qubit Hamiltonian} and discussion below).\\
\indent The confinement potential, created by electrical gating of the 2DEG, at each DQD is modeled as a quartic potential centered at $(\pm R,0)$\cite{Burkard1999,Stepanenko2007}:
\begin{equation}\label{Quartic_potential}
V_{\pm R}(x,z)=\frac{m\omega_0^2}{2}\left\lbrace\frac{1}{4a^2}\left[(x\mp R)^2
-a^2\right]^2+z^2\right\rbrace,
\end{equation}
where $m$ is the effective mass of the electron and $\hbar \omega_0$ is the confinement energy. The two local minima of the quartic potential can be treated as isolated harmonic wells with ground-state wave functions $\phi_{\pm a}$. The single-particle single-dot states are orthonormalized in order to define a basis for the two-electron states of the DQD\cite{Burkard1999}:
\begin{equation}\label{orthonormalized_single_particle_state}
\psi_{\pm a}=\frac{\phi_{\pm a}-g\phi_{\mp a}}{\sqrt{1-2s g +g^2}},
\end{equation}
where $s=\langle\phi_a\vert\phi_{-a}\rangle=\exp[-a^2/a_B^2]$ is the overlap of the harmonic ground-state wave functions, $a_B=\sqrt{\hbar/m\omega_0}$ is the effective Bohr radius of a single harmonic well, and $g=(1-\sqrt{1-s^2})/s$ is a mixing factor. Including doubly occupied states, the two-electron basis states are
\begin{align}
\vert S(2,0)\rangle&=\psi_{-a}\psi_{-a}, \\
\vert S(0,2)\rangle&=\psi_a\psi_a,\\
\vert S(1,1)\rangle&=\frac{1}{\sqrt{2}}(\psi_{-a}\psi_a +\psi_a\psi_{-a}),\\
\vert T_0\rangle&=\frac{1}{\sqrt{2}}(\psi_{-a}\psi_a -\psi_a\psi_{-a}),
\end{align}
where $(n_L,n_R)$ represents the number of electrons in the left and right dots of the DQD.\\
\indent The bias $\varepsilon$ is modeled as an energy difference between single-dot ground orbitals in a DQD \cite{Stepanenko2007}, which can be pictured as the tilting of the quartic potential, as seen in Fig. \ref{fig:1}. Such difference is created by an additional bias potential $V_b$ generated by controllable electrostatic gates \cite{Petta2005} (see Appendix \ref{app:Single-qubit Hamiltonian}). At $\varepsilon \approx 0$, when the ground orbitals of each dot have the same energy, the lowest energy state of two electrons is the singlet with singly occupied orbitals $\vert S(1,1)\rangle$. Whereas for $\vert\varepsilon\vert\gg U$, where $U$ is the on-site Coulomb repulsion, the doubly occupied singlet state ($\vert S(0,2)\rangle$ or $\vert S(2,0)\rangle$, depending on the tilting direction of the potential) is the ground state of the two electrons. In the intermediate region, where $\varepsilon \approx U$, the ground state is a hybridization of the singly and doubly occupied states  (unless stated otherwise, hereinafter we will consider $\vert S(0,2)\rangle$ as the doubly-occupied ground state) expressed as \cite{Taylor2005,Taylor2007}
\begin{equation}\label{hybsinglet}
\vert \tilde{S}\rangle =\sin\theta\vert S(0,2)\rangle+\cos\theta\vert S(1,1)\rangle,
\end{equation}
where $\theta$ is a mixing angle that parametrizes the hybridization (Appendix \ref{app:Single-qubit Hamiltonian}).\\
\indent In the ground state, by tuning the bias $\varepsilon$, the two electrons in each DQD can undergo a transition from $S(0,2)$ to $S(1,1)$ and vice versa, while the spin-symmetric triplet state $\vert T_0\rangle$ is cornered in the (1,1) charge configuration due to Pauli exclusion. This change in the charge configuration along with the state-dependent Coulomb coupling between electron pairs creates an entangling interaction between qubits \cite{Taylor2005,Stepanenko2007,Shulman2012}. In this light, the interaction Hamiltonian of the two-qubit system in the state basis $\lbrace \vert  \tilde{S}\tilde{S}\rangle,\vert \tilde{S}T_0\rangle,\vert T_0 \tilde{S}\rangle,\vert T_0 T_0\rangle\rbrace$ is $H_{int}=diag\lbrace
V_{\tilde{S}\tilde{S}}, V_{\tilde{S}T_0},V_{T_0 \tilde{S}},V_{T_0T_0}\rbrace$, where $V_{ij}$ are the Coulomb integrals involving the hybridized singlet \eqref{hybsinglet} and the unpolarized triplet states (their closed-form expressions are given in Appendixes \ref{app:Interaction_Hamiltonian} and \ref{app:Calculation_Coulomb_integrals}). Expressing $H_{int}$ in terms of the $SU(4)$ generators, up to an identity term, gives \cite{Ramon2011}
\begin{equation}\label{interactionHamiltonian}
H_{int}=-\beta_1(\varepsilon_1,\varepsilon_2)\sigma_z\otimes I -\beta_2(\varepsilon_1,\varepsilon_2) I\otimes\sigma_z + \alpha(\varepsilon_1,\varepsilon_2)\sigma_z\otimes\sigma_z,
\end{equation}
where the subscript 1 (2) corresponds the left (right) qubit and
\begin{align}
\beta_1(\varepsilon_1,\varepsilon_2)&=\frac{1}{4}\left(-V_{\tilde{S}\tilde{S}}-V_{\tilde{S}T_0}+V_{T_0 \tilde{S}}+V_{T_0T_0}\right),\label{beta_1}\\
\beta_2(\varepsilon_1,\varepsilon_2)&=\frac{1}{4}\left(-V_{\tilde{S}\tilde{S}}+V_{\tilde{S}T_0}-V_{T_0 \tilde{S}}+V_{T_0T_0}\right),\label{beta_2}\\
\alpha(\varepsilon_1,\varepsilon_2)&=\frac{1}{4}\left(V_{\tilde{S}\tilde{S}}-V_{\tilde{S}T_0}-V_{T_0 \tilde{S}}+V_{T_0T_0}\right),\label{alpha}
\end{align}
are the local ($\beta_1$, $\beta_2$) and nonlocal ($\alpha$) terms of the interaction Hamiltonian.\\
\indent Each qubit is separately controlled by tuning the bias $\varepsilon$ and, consequently, changing the strength of the energy splitting, $J(\varepsilon)$, between the singlet and triplet states. The DQD is also under the influence of an external magnetic field that shifts the energy of the polarized triplet states making them energetically inaccessible. This external field is assumed parallel to the plane of the 2DEG\cite{Shulman2012} and not strong enough to affect the exchange energy $J(\varepsilon)$ (the wave function compression caused by the external magnetic field is perpendicular to the 2DEG's plane, where the confinement potential is already strong, see Appendix \ref{app:Single-qubit Hamiltonian}). At the same time, each spatially separated electron in the DQD experiences a slightly different effective magnetic field due to hyperfine interaction with the inhomogeneous bath of nuclear spins, leading to a magnetic field gradient across the DQD with an energy $h$. As a result, the single-qubit Hamiltonian $H_i$, $i=1,2$, in the basis $\lbrace \vert \tilde{S}\rangle,\vert T_0\rangle\rbrace$, and assuming $\cos\theta >\sin\theta$ in Eq. \eqref{hybsinglet}, is given by
\begin{equation}\label{singleHamiltonian}
H_i=\frac{1}{2}
\begin{pmatrix}
J_i & h_i \\
h_i & -J_i
\end{pmatrix}.
\end{equation}
\indent Consider now the whole system. The two-qubit Hamiltonian, which combines both single-qubit and interaction Hamiltonians, is expressed in terms of the Pauli operators $\sigma_x=\vert\tilde{S}\rangle\langle T_0\vert + \vert T_0\rangle\langle\tilde{S}\vert$, $\sigma_z=\vert\tilde{S}\rangle\langle\tilde{S}\vert - \vert T_0\rangle\langle T_0\vert$, $I=\vert\tilde{S}\rangle\langle\tilde{S}\vert +\vert T_0\rangle\langle T_0\vert $, (with $\cos\theta >\sin\theta$ in Eq. \eqref{hybsinglet} for both qubits) as
\begin{widetext}
\begin{equation}\label{Two-qubit_Hamiltonian}
H=\left(\frac{J_1(\varepsilon_1)}{2}-\beta_1(\varepsilon_1,\varepsilon_2)\right)\sigma_z\otimes I + \frac{h_1}{2}\sigma_x\otimes I + \left(\frac{J_2(\varepsilon_2)}{2}-\beta_2(\varepsilon_1,\varepsilon_2)\right)I\otimes\sigma_z + \frac{h_2}{2}I\otimes\sigma_x + \alpha\left(\varepsilon_1,\varepsilon_2\right)\sigma_z\otimes\sigma_z.
\end{equation}
\end{widetext}

\indent For the analysis and results reported in this work, we chose our model's parameters to be similar to the experimental values reported in the literature for capacitively coupled qubits \cite{Shulman2012}. Accordingly, we have modeled the double dots using a confinement energy $\hbar\omega_0=1 \mathrm{meV}$ ($a_B\approx 33.7\mathrm{nm}$), the effective electron mass for GaAs  $m=0.067m_e$ ($m_e$ is the electron mass), a half interdot distance $a=2.5a_B$ and a half interqubit distance (from each DQD's center) $R=7.5a_B$. The biases in the Hamiltonian \eqref{Two-qubit_Hamiltonian} are restricted to have a maximum of $\varepsilon_{max}\approx3.344\mathrm{meV}$, resulting in maximum parameter values $J_i(\varepsilon_{max})\approx 1.26 \mathrm{\mu eV}$ and $\alpha(\varepsilon_{max},\varepsilon_{max})\approx 3.81 \mathrm{neV}$, in good agreement with the experimentally reported values of $J_i(\varepsilon_{max}) \approx 2\pi\times 300 \mathrm{MHz}$ and $\alpha(\varepsilon_{max},\varepsilon_{max})\approx 2\pi\times 0.87 \mathrm{MHz}$ \cite{Shulman2012}. (This bias corresponds to a ratio between singly and doubly occupied singlet content in $\vert \tilde{S}\rangle$ of about 0.13.) The average magnetic field gradient energy is fixed as $h_i = J_i(\varepsilon_{max})/10$, again reflecting the experimentally reported value of $h_i \approx 2\pi\times 30 \mathrm{MHz}$ \cite{Shulman2012}.

\section{Local invariants of two-qubit unitary gates}\label{sec:Local_invariants}
All two-qubit gates belong to the unitary group $SU(4)$, which is comprised of a set of local operations $SU(2)\otimes SU(2)$ and a set of nonlocal operations $SU(4)\backslash SU(2)\otimes SU(2)$. The latter is further divided into a set of perfect entangling operations, i.e., those that can generate maximally entangled states (e.g., CNOT), and the set of nonlocal operations that are not perfect entanglers (e.g., SWAP). This division of the group $SU(4)$ is a consequence of the Cartan decomposition of its Lie algebra \cite{Khaneja2001a,Kraus2001,Zhang2003}, which enables us to express any $U \in SU(4)$ as
\begin{equation}
U = k_1Ak_2
  = k_1\exp\left\lbrace -\frac{i}{2}\underset{j=x,y,z}{\sum}\gamma_j\sigma_j\otimes\sigma_j \right\rbrace k_2,
\end{equation}
where $k_i\in SU(2)\otimes SU(2)$ are local operations and $\mathsf{span}\frac{i}{2}\left\lbrace\sigma_x\otimes\sigma_x,\sigma_y\otimes\sigma_y,\sigma_z\otimes\sigma_z\right\rbrace$ generates the maximal Abelian subalgebra (of the algebra spanned by $\sigma_i\otimes\sigma_j,i,j=x,y,z$), which in turn generates $SU(4)\backslash SU(2)\otimes SU(2)$. Because of the commuting nature of the two-qubit operators in $A$'s exponent, the nonlocal content and, therefore, the entangling power of the gate $U$ can be completely represented by the three real numbers $\lbrace \gamma_x,\gamma_y,\gamma_z\rbrace$.\\
\indent The above representation of the nonlocal content of a two-qubit gate is not unique. In Ref. \onlinecite{Makhlin2002}, the unitary and symmetric matrix $m(U)$, defined as $m(U)=(Q^{\dagger}UQ)^TQ^{\dagger}UQ$, is introduced. Here, $Q$ denotes the transformation of the matrix $U$ from the logical basis into the Bell basis. In this basis any local operation is represented by a real orthogonal matrix, i.e., $Q^{\dagger}k_iQ\in SO(4)$, making the spectrum of $m$ invariant under local operations. In fact, the terms that form the characteristic polynomial of the matrix $m(U)$, and thus determine its spectrum, give a complete set of \textit{local invariants} \cite{Makhlin2002}:
\begin{equation}\label{local_invariants}
\begin{aligned}
G_1 & =\mathrm{Re}\left[\frac{\mathrm{tr}^2[m(U)]}{16 }\right],\\
G_2 & =\mathrm{Im}\left[\frac{\mathrm{tr}^2[m(U)]}{16 }\right],\\
G_3 & =\frac{\mathrm{tr}^2[m(U)]-\mathrm{tr}[m^2(U)]}{4}.
\end{aligned}
\end{equation}
These local invariants convey the nonlocal properties of the matrix $U$ and have a one-to-one correspondence with the parameters $\lbrace \gamma_x,\gamma_y,\gamma_z\rbrace$ \cite{Zhang2003}.\\
\indent The set of numbers $\lbrace \gamma_x,\gamma_y,\gamma_z\rbrace$, in which the two-qubit gate $U$ is periodic, have a geometric structure of a 3-torus. An equivalent representation of the points on the 3-torus is a cube with side length equal to $\pi$. Each point in this cube corresponds to a nonlocal two-qubit operation, but different points may represent the same two-qubit gate up to local transformations. This symmetry can be reduced using the Weyl reflection group\cite{Zhang2003}, which is generated by permutations of all elements or permutations with sign flips of two elements of $\lbrace \gamma_x,\gamma_y,\gamma_z\rbrace$. The symmetry reduction leads to a  Weyl chamber, in this case a tetrahedron delimited by $(\gamma_x,\gamma_y,\gamma_z)=(0,0,0),(\pi,0,0),(\pi/2,\pi/2,0),(\pi/2,\pi/2,\pi/2)$, as shown in Fig. \ref{fig:3b}. Each point in the Weyl chamber is a unique representation of a class of two-qubit gates that are equivalent up to local operations (except on the base where a local equivalence class may be represented by two symmetric equivalent points \cite{Zhang2003}). For example, a CNOT gate and a CPHASE gate are equivalent up to local operations and are represented by the same point $(\pi/2,0,0)$ in the Weyl chamber (see Table \ref{tab:two-qubit_gates}). Therefore, the Weyl chamber is a geometric representation of all possible two-qubit gates. \\
\begin{table}[tbp]
\caption{Some well-known two-qubit gates along with their local invariants $G_1$, $G_2$, $G_3$, and their coordinates ($\gamma_x,\gamma_y,\gamma_z$) in the Weyl chamber. Among them, only I (identity) and SWAP are not perfect entanglers.}
\begin{tabular}{l||c c c c c c}
Gate & $G_1$ & $G_2$ & $G_3$ & $\gamma_x$ & $\gamma_y$ & $\gamma_x$ \\
\hline
\hline
I & 1 & 0 & 3 & 0 & 0 & 0\\
\hline
SWAP & -1 & 0 & -3 & $\pi/2$ & $\pi/2$ & $\pi/2$\\
\hline
CNOT & 0 & 0 & 1 & $\pi/2$ & 0 & 0\\
\hline
CPHASE & 0 & 0 & 1 & $\pi/2$ & 0 & 0\\
\hline
iSWAP & 0 & 0 & -1 & $\pi/2$ & $\pi/2$ & 0\\
\hline
$\sqrt{\mathrm{SWAP}} $ & 0 & 1/4 & 0 & $\pi/4$ & $\pi/4$ & $\pi/4$\\
\end{tabular}
\label{tab:two-qubit_gates}
\end{table}
\indent The nonlocal properties of a Hamiltonian, specifically its capability to generate perfect entangling operations, follow a condition stated by Makhlin \cite{Makhlin2002}: \textit{a two-qubit gate $U$ is a perfect entangler if and only if the convex hull of the eigenvalues of $m(U)$ contains zero}. This  condition, in terms of the local invariants, reads as \cite{Makhlin2002}
\begin{equation}\label{Makhlin_perfect_entangler_condition}
\begin{aligned}
&\sin^2 \varphi\leq 4\vert G\vert\leq 1\\
\mathrm{and}\\
&\cos \varphi(\cos\varphi - G_3)\geq 0,
\end{aligned}
\end{equation}
where $G=G_1+iG_2=\vert G\vert e^{i\varphi}$. Similarly, in Ref. \onlinecite{Zhang2003} it is shown that a two-qubit gate is a perfect entangler as long as $\gamma_x$, $\gamma_y$, $\gamma_z$ fulfill either of the following two conditions:
\begin{equation}\label{Zhang_perfect_entangler_condition}
\begin{aligned}
&\frac{\pi}{2}\leq \gamma_i+\gamma_k\leq \gamma_i +\gamma_j +\frac{\pi}{2}\leq \pi\\
\mathrm{or}\\
&\frac{3\pi}{2}\leq \gamma_i+\gamma_k\leq \gamma_i +\gamma_j +\frac{\pi}{2}\leq 2\pi,
\end{aligned}
\end{equation}
where $(i,j,k)$ is a permutation of $(x,y,z)$. This corresponds to a polyhedron in the Weyl chamber delimited by the points $(\gamma_x,\gamma_y,\gamma_z)=
(\pi/2,0,0),(\pi/2,\pi/2,0),(\pi/4,\pi/4,0),
(\pi/4,\pi/4,\pi/4)$,
$(3\pi/4,\pi/4,0),
(3\pi/4,\pi/4,\pi/4)$, (Fig. \ref{fig:3b}). Both conditions for the generation of perfect entangling gates, Eqs.  \eqref{Makhlin_perfect_entangler_condition} and \eqref{Zhang_perfect_entangler_condition}, are equivalent.\\
\begin{figure}[tbp]
    \centering
    \subfigure{
    \includegraphics[trim=0cm 7.5cm 0cm 7.5cm, clip=true,width=8.5cm, angle=0]{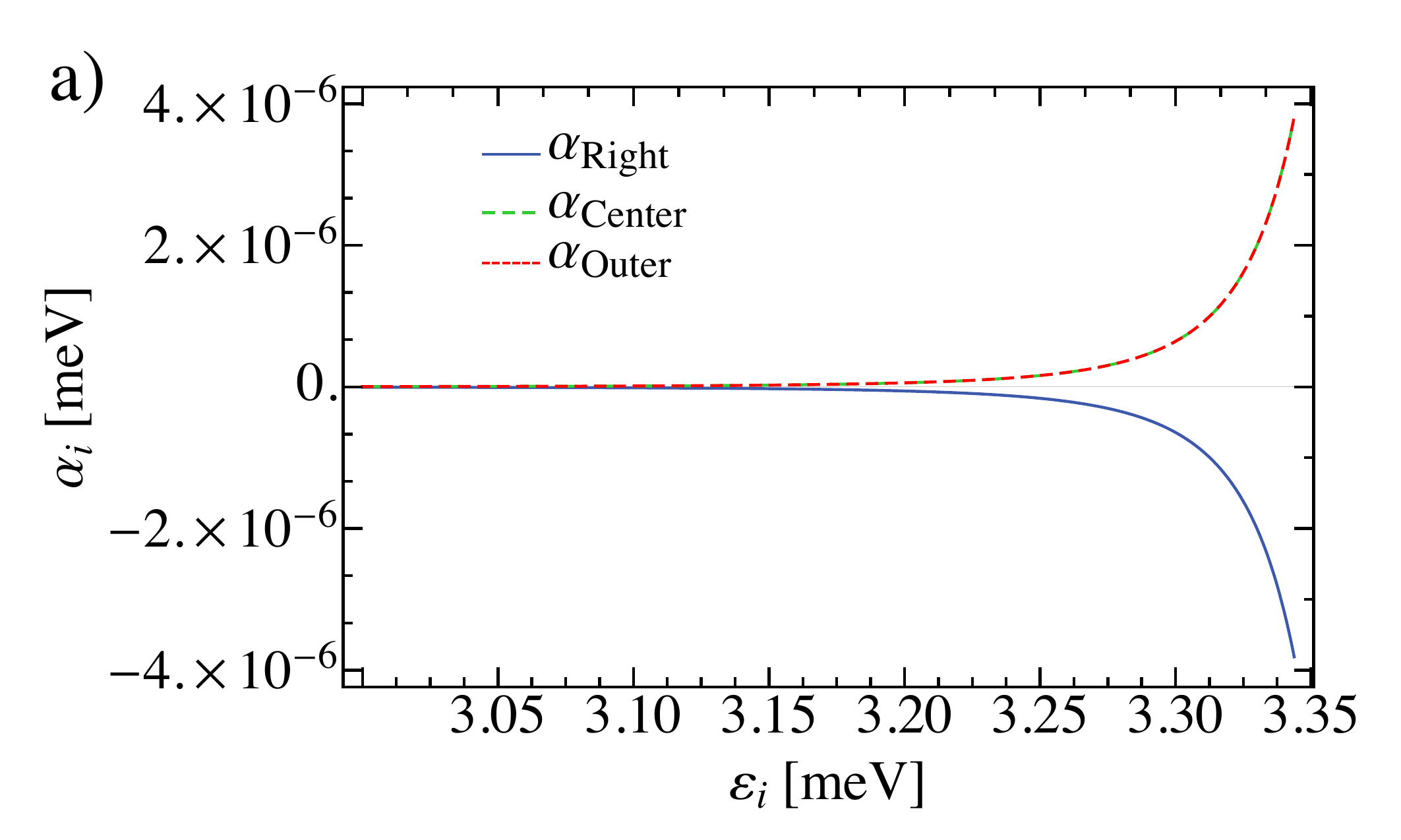}
    \label{fig:2a}}\\
    \subfigure{
    \includegraphics[trim=0cm 7.5cm 0cm 7.5cm, clip=true,width=8.5cm, angle=0]{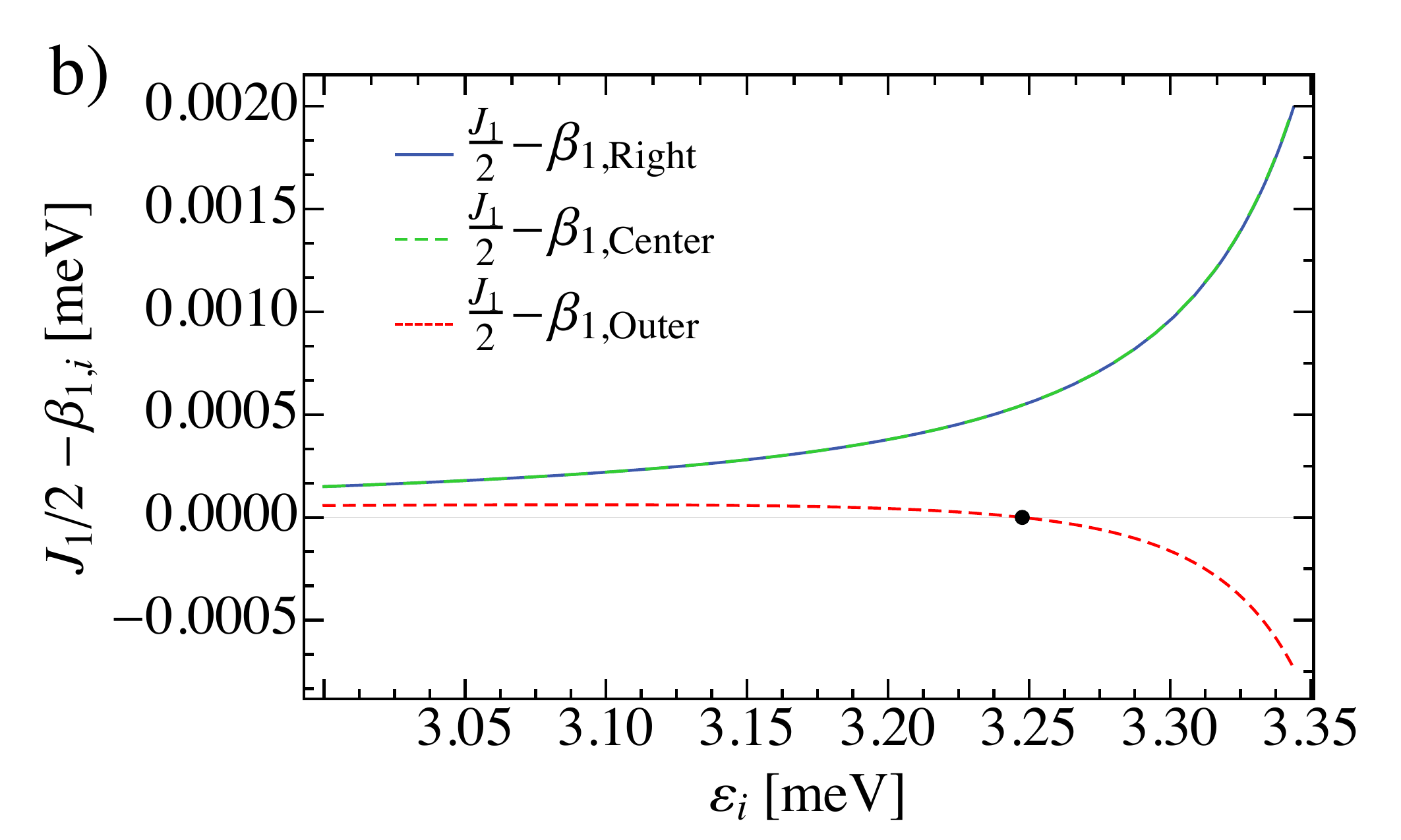}
    \label{fig:2b}}\\
    \subfigure{
    \includegraphics[trim=0cm 7.5cm 0cm 7.5cm, clip=true,width=8.5cm, angle=0]{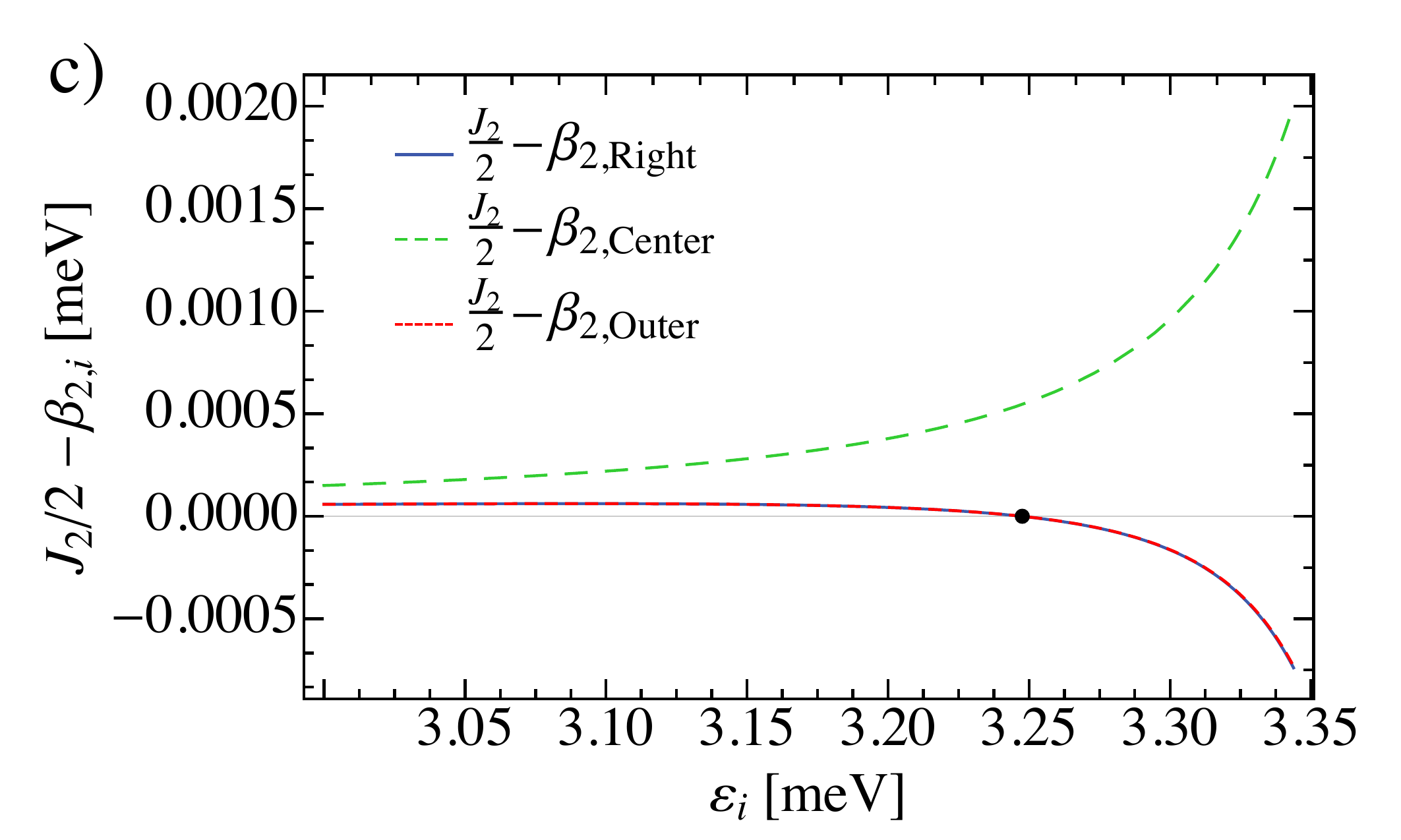}
     \label{fig:2c}}
    \caption{(Color online.) Hamiltonian parameters vs control bias $\varepsilon_i$, where the same magnitude of bias is applied on each qubit. Results are shown for each of the three ways to apply the bias. \textbf{(a)} Nonlocal (coupling) term. On this scale, the \textit{Right} and \textit{Center} cases are indistinguishable. \textbf{(b)} Local exchange term of the left qubit. \textit{Right} and \textit{Center} cases are again indistinguishable. \textbf{(c)} Local exchange term of the right qubit. \textit{Right} and \textit{Outer} cases are indistinguishable.  Local exchange terms vanish at $\varepsilon_1=\varepsilon_2\approx 3.25\mathrm{meV}$ (represented by a black dot).}
\end{figure}

\section{Biasing modes and perfect entangling gates}\label{sec:Charge_configuration_entangling_gates}
\indent We are interested in finding the perfect entangling gates that can be directly generated by the evolution of the two-qubit system. We know that the capacitive coupling depends on the charge configuration of both DQDs, where the Coulomb matrix elements of the interaction Hamiltonian, Eq. \eqref{interactionHamiltonian}, are sensitive to the way in which the confinement potential in each DQD is biased. In this light, we investigate three possible biasing modes:
\begin{itemize}
\item \textit{Right}: The confinement potentials, of both DQDs, are tilted to the same side (either side is equivalent, we choose the right side), leading to  an increase of the $\vert S(0,2)\rangle$-content in the hybridized singlet state of both qubits.
\item \textit{Center}: The confinement potentials are tilted towards the two-DQD center, causing the increase of the $\vert S(0,2)\rangle$-content on the left qubit and the $\vert S(2,0)\rangle$-content on the right qubit.
\item \textit{Outer}: The confinement potentials are tilted away from the two-DQD center, with a consequent increase of the $\vert S(2,0)\rangle$-content on the left qubit and the $\vert S(0,2)\rangle$-content on the right qubit.
\end{itemize}

\begin{figure}[tbp]
    \centering
    \subfigure{
    \includegraphics[trim=0cm 6cm 0cm 6cm, clip=true,width=8.5cm, angle=0]{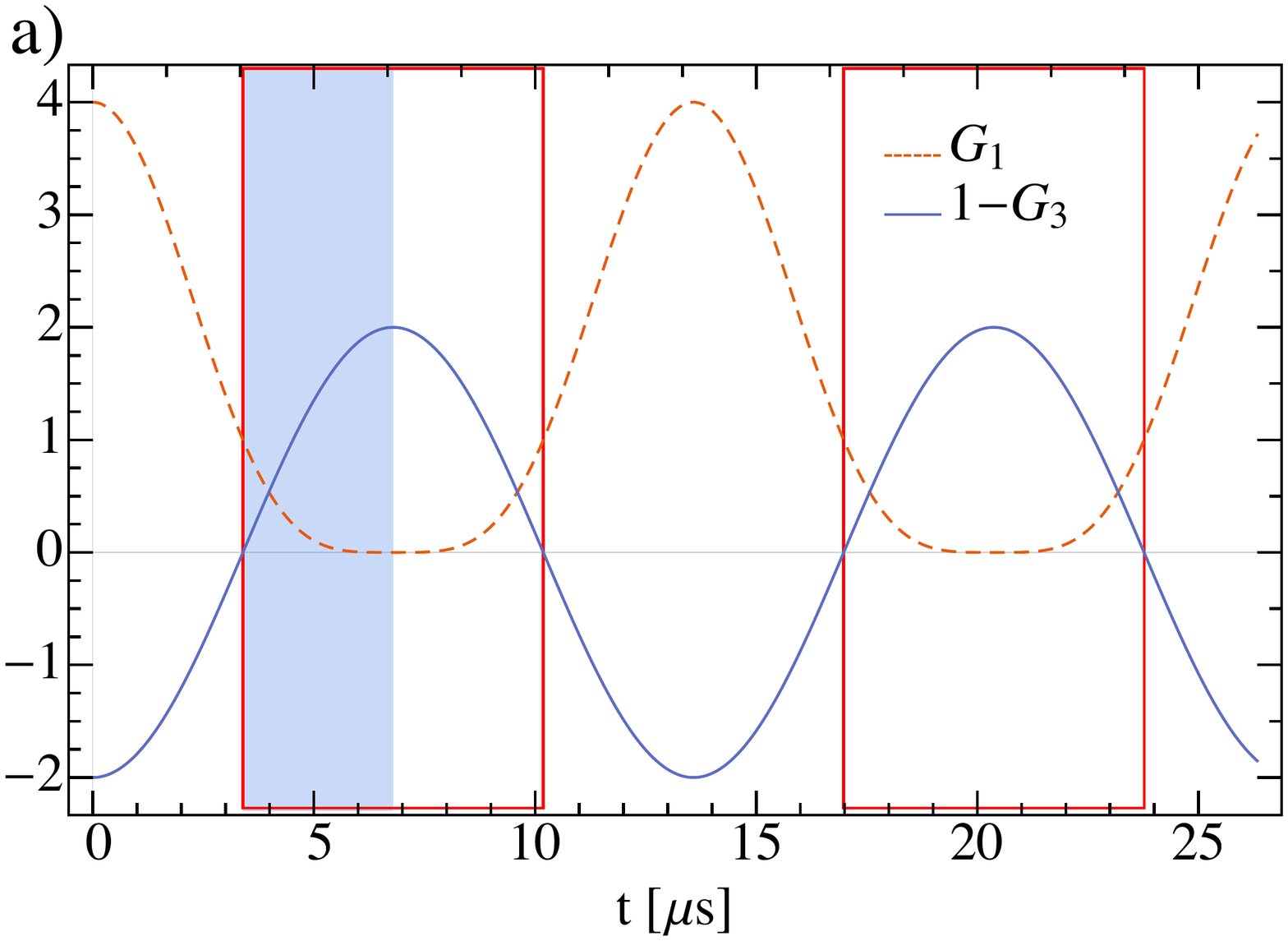}
    \label{fig:3a}}\\
    \subfigure{
    \includegraphics[trim=0cm 8cm 0cm 8cm, clip=true,width=8.6cm, angle=0]{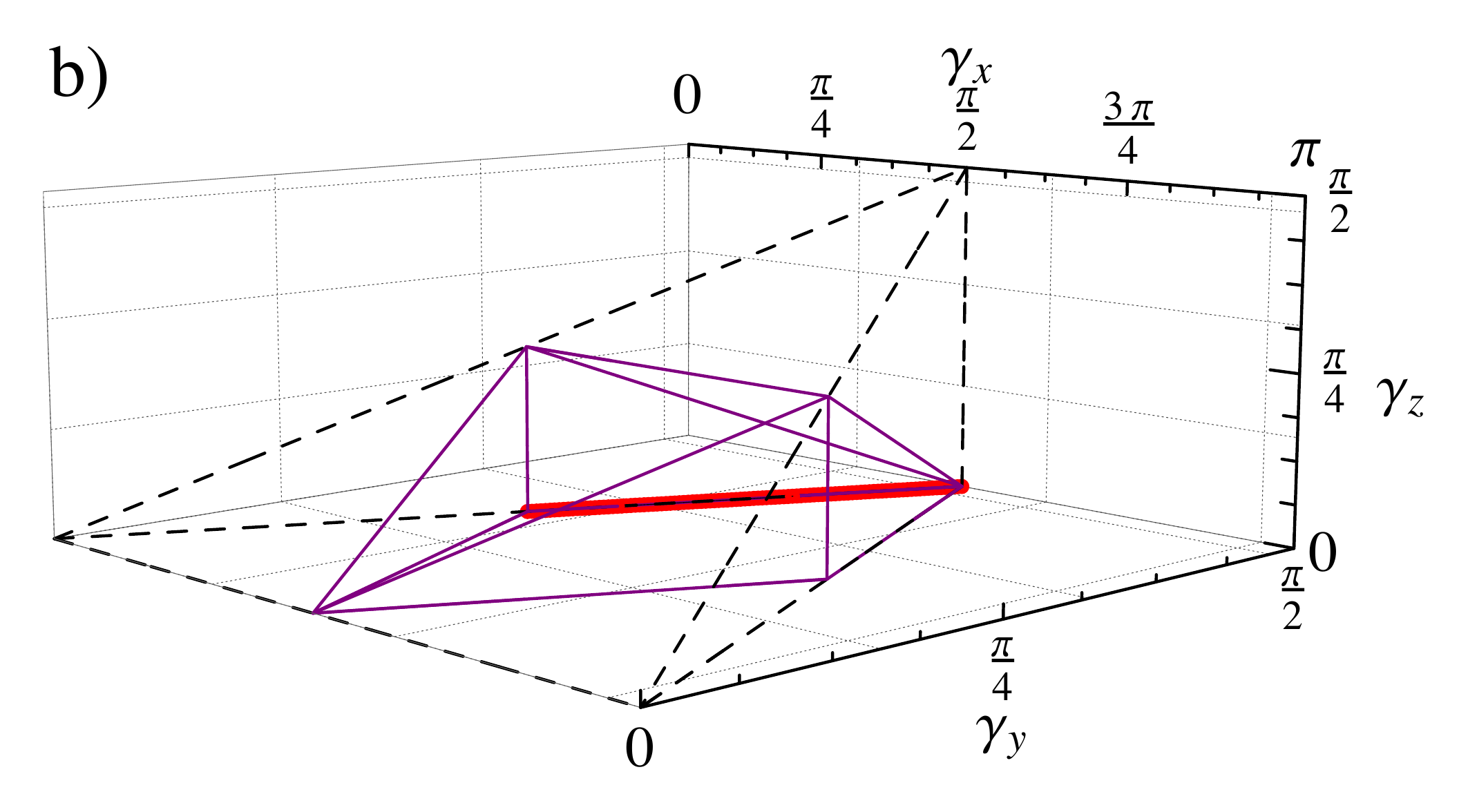}
    \label{fig:3b}}
    \caption{(Color online.) \textbf{(a)} Invariants $G_1$ and $1-G_3$. A set of perfect entangling gates are generated for times enclosed by red rectangles. \textbf{(b)} The tetrahedral Weyl chamber (black dashed lines), geometric representation of all possible two-qubit gates, comprises non-entangling and entangling gates, the latter being all enclosed by the polyhedral volume delimited with purple solid lines. On the edge of this polyhedron lies the trajectory (red thick line) of the evolution operator for the shaded time interval of panel \textbf{(a)}.}
\end{figure}

\begin{figure}[tbp]
    \centering
    \subfigure{
    \includegraphics[trim=0cm 7.3cm 0cm 7.3cm, clip=true,width=8.5cm, angle=0]{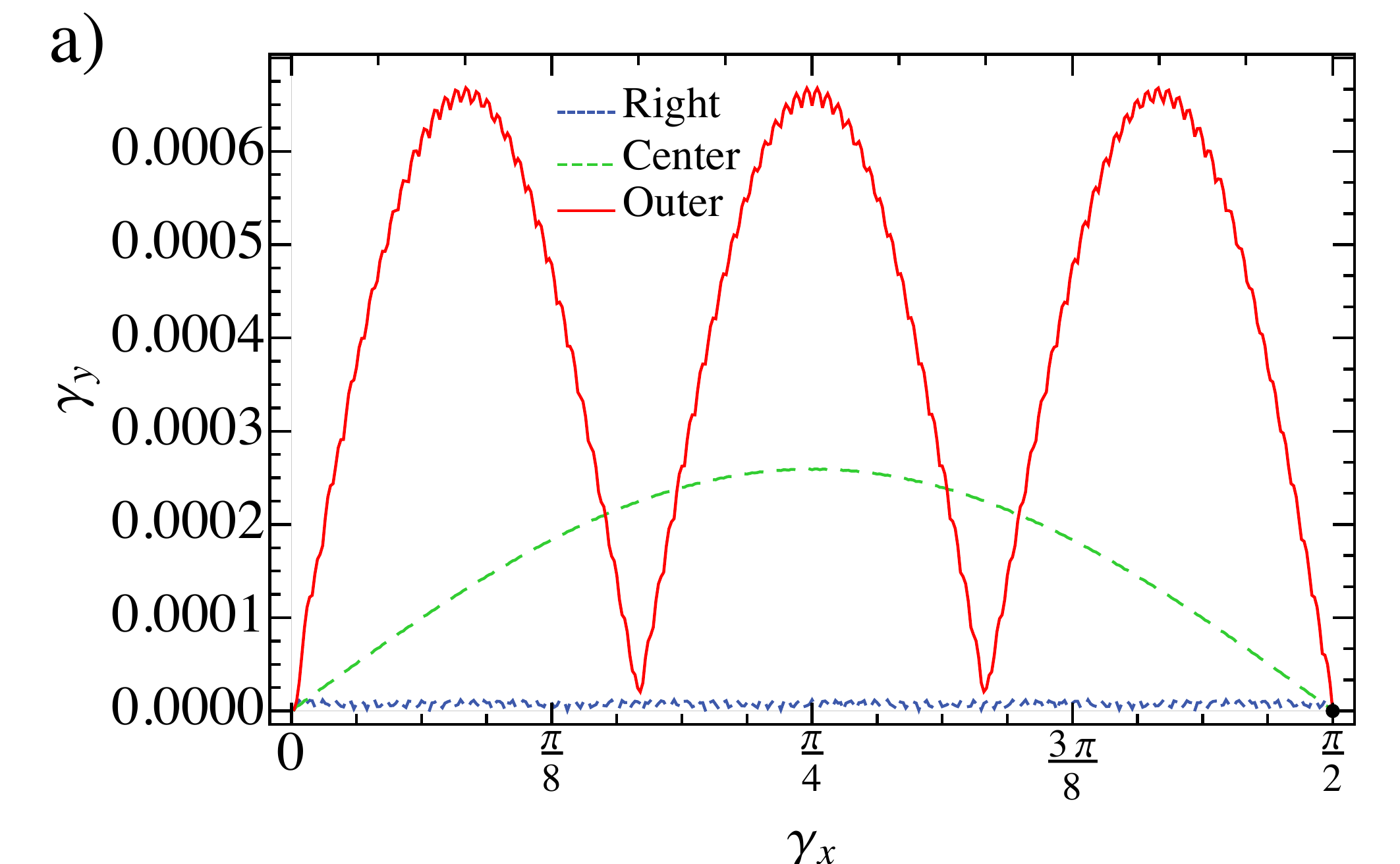}
    \label{fig:4a}}\\
    \subfigure{
    \includegraphics[trim=0cm 8cm 0cm 8cm, clip=true,width=8.6cm, angle=0]{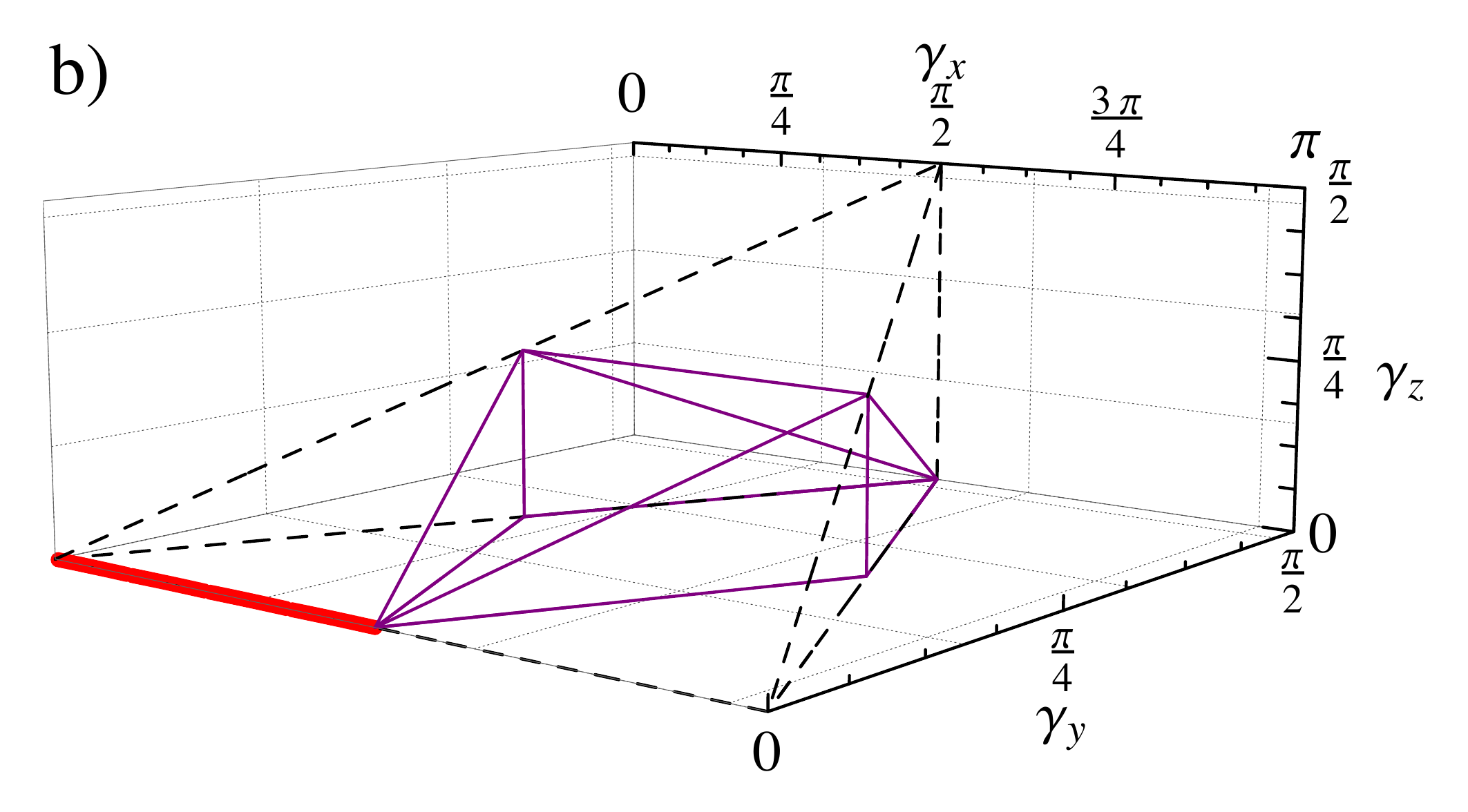}
    \label{fig:4b}}
    \caption{(Color online.) \textbf{(a)} Best-time Weyl chamber trajectories, projected on the $\gamma_x-\gamma_y$ plane, that reach the CNOT gate $(\pi/2,0,0)$ for each of the three biasing modes: \textit{Right}, \textit{Center} and \textit{Outer}, respectively. \textbf{(b)} Weyl chamber trajectory (thick red line) during generation of the CNOT.}
\end{figure}

\indent As seen in Fig. \ref{fig:2a}, when both qubits are tilted to the same side (in this case, to the right) the sign of the nonlocal term $\alpha$ (Eq. \eqref{alpha}) is negative, whereas for a tilting toward or away from the two-DQD center, $\alpha$ is positive. Nevertheless, the magnitude of $\alpha$ in the three scenarios is the same ($\vert\alpha(\varepsilon_{max},\varepsilon_{max})\vert\approx 3.81\mathrm{neV}$). However, the magnitude and sign of the local terms ($J_i/2 -\beta_i,i=1,2$), Eqs. \eqref{beta_1} and \eqref{beta_2}, varies in each case (see Figs.\ref{fig:2b} and \ref{fig:2c}).\\
\indent Of particular interest is the case where both qubits are tilted away from the two-DQD center. Notice that at a certain point ($\varepsilon_1=\varepsilon_2\approx 3.25\mathrm{meV}$), with both qubits being biased symmetrically, the local exchange terms of both qubits vanish, leaving a simpler Hamiltonian
\begin{equation}\label{Hamiltonian_z_vanished}
H_0=\frac{h}{2}\sigma_x\otimes I + \frac{h}{2}I\otimes\sigma_x + \alpha(\varepsilon_1,\varepsilon_2)\sigma_z\otimes\sigma_z,
\end{equation}
with $\alpha\approx 0.152 \mathrm{neV}$ and $h\approx 0.126\mathrm{\mu eV}$ (hereafter we make the reasonable assumption that the average magnetic field gradient is the same in both qubits, $h_1=h_2=h$).\\ \indent The advantage of using certain bias positions to simplify  the Hamiltonian was first pointed out in Ref.~\onlinecite{Ramon2011}. That work treats the case of large qubit coupling, $\alpha\gg h$ (which implies that the hybrid state has mostly doubly occupied singlet content as in Ref.~\onlinecite{VanWeperen2011}), and the bias is tuned to the point where both qubits satisfy $J_i=2(\beta_i-\alpha)$ to obtain a CPHASE gate. In contrast, we consider the case where the hybrid state, $\vert \tilde{S}\rangle$, is close to the singly occupied singlet state, $\vert S(1,1)\rangle$, leading to longer coherence times but smaller coupling, $\alpha\ll h$, similar to the experimental situation \cite{Shulman2012}. This leads to a more complicated evolution since the off-diagonal magnetic field gradient terms of the Hamiltonian are no longer negligible. However, we will show that a maximally entangling gate can nonetheless be generated directly.\\
\indent In order to characterize the type of entangling gates, up to local operations, that can be directly generated by the Hamiltonian $H_0$ \eqref{Hamiltonian_z_vanished}, we derive the
set of local invariants \eqref{local_invariants} for the operator $U=\exp\left[-i t H_0\right]$:
\begin{equation}
\begin{aligned}
G_1&=\frac{\left(h^2+(h^2+\alpha^2)\cos(2t\alpha)+\alpha^2\cos\left(2t\sqrt{h^2+\alpha^2}\right)\right)^2}{4\left(h^2+\alpha^2\right)^2},\\
G_2&=0,\\
G_3&=1+\frac{2\cos(2t\alpha)\left(h^2+\alpha^2\cos\left(2t\sqrt{h^2+\alpha^2}\right)\right)}{h^2+\alpha^2}.
\end{aligned}
\end{equation}
With these local invariants the conditions for the generation of perfect entangling gates, Eq. \eqref{Makhlin_perfect_entangler_condition}, turn into
\begin{equation}\label{inequalities_for_H0}
\begin{aligned}
&0\leq \frac{\left(h^2+(h^2+\alpha^2)\cos(2t\alpha)+\alpha^2\cos\left(2t\sqrt{h^2+\alpha^2}\right)\right)^2}{\left(h^2+\alpha^2\right)^2} \leq 1\\
&\mathrm{and}\\
&-\frac{2\cos(2t\alpha)\left(h^2+\alpha^2\cos\left(2t\sqrt{h^2+\alpha^2}\right)\right)}{h^2+\alpha^2} \geq 0.
\end{aligned}
\end{equation}
\indent By plotting these inequalities in Fig. \ref{fig:3a}, we find that the two-qubit operation $U=\exp\left[-i t H_0\right]$ generates a set of perfect entangling gates within the time interval enclosed in the rectangles, $\left[\frac{\pi+4\pi n}{4\alpha},\frac{4\pi(n+1)-\pi}{4\alpha}\right]$, with $n=0,1,2,\ldots$. The interval is symmetric around $\frac{\pi(2n+1)}{2\alpha}$, i.e., the gates generated in the shaded interval (Fig. \ref{fig:3a}), $\left[\frac{\pi+4\pi n}{4\alpha},\frac{\pi(2n+1)}{2\alpha}\right]$, are equivalent to the gates generated in the unshaded half, $\left[\frac{\pi(2n+1)}{2\alpha},\frac{4\pi(n+1)-\pi}{4\alpha}\right]$. The coordinates ($\gamma_x,\gamma_y,\gamma_z$) in the Weyl chamber that correspond to those perfect entangling gates are depicted in Fig. \ref{fig:3b}. The tetrahedral Weyl chamber is depicted with black dashed lines, while the solid polyhedron, delimited with purple solid lines, represents the volume that encloses all perfect entangling gates \cite{Zhang2003}. The red thick line, on the edge of the aforementioned polyhedral volume, represents the Weyl chamber trajectory of the operation $U=\exp\left[-i t H_0\right]$ in the time interval $\left[\frac{\pi}{4\alpha},\frac{\pi}{2\alpha} \right]$. Among the perfect entangling gates that can be generated, up to local operations, with the Hamiltonian $H_0$ is the iSWAP gate \cite{Schuch2003} (locally equivalent to the DCNOT gate \cite{Zhang2004a}, which is formed by implementing a CNOT operation from the first qubit onto the second qubit and then a second CNOT operation from the second qubit onto the first qubit \cite{Collins2001}), with coordinates ($\pi/2,\pi/2,0$) (see Table \ref{tab:two-qubit_gates}). Due to the weak-coupling value taken here, the time required to generate the iSWAP gate is rather long, $t\approx 6.8\mathrm{\mu s}$, but still within the coherence time range for GaAs qubits reported in the literature \cite{Bluhm2010}.\\
\indent We also investigate the type of gates that can be generated with a single implementation of the evolution operator using the complete two-qubit Hamiltonian $H$, Eq.\eqref{Two-qubit_Hamiltonian}. The expressions of the local invariants corresponding to $U=\exp[-i t H]$ involve nested summations and the roots of a quartic polynomial, which make it rather difficult to obtain any general insight into the type of entangling gates that can be generated directly. Instead, we target well-known perfect entangling gates (Table \ref{tab:two-qubit_gates}) and search numerically for any realization of $U$ that is locally equivalent to the desired entangling gate. Accordingly, using the local invariants equations \eqref{local_invariants} for the evolution operator, $U$, and for a given familiar entangling gate, $\mathcal{U}$, we define an objective function
\begin{equation}\label{Objective_function}
f=\overset{3}{\underset{i=1}{\sum}}\Delta G_i^2, \ \mathrm{with} \ \Delta G_i=\vert G_i(U)-G_i(\mathcal{U})\vert.
\end{equation}
We used the numerical optimization method \textit{differential evolution} embedded in Mathematica$^\circledR$'s \textit{NMinimize} function to minimize the objective function \eqref{Objective_function} under the constraint $0<\varepsilon_i\leq\varepsilon_{max}$ and $t_{min}<t<t_{min}+40\mathrm{ns}$, where $t_{min}=\frac{\pi}{4\vert\alpha(\varepsilon_{max},\varepsilon_{max})\vert}$. The parameters differential-weight$=1$ and crossover-probability$=0.5$, within the differential evolution algorithm, yielded the best performance in our particular case. We searched over 200 random seeds for solutions that minimize the objective function with a tolerance equal to $10^{-6}$ (i.e., any solution that yields a value in the objective function greater than $10^{-6}$ is discarded). Using more random seeds could result in additional solutions being found, but would also increase the runtime. The solutions found were further optimized using the function \textit{FindRoot} in Mathematica$^\circledR$, with a tolerance equal to $10^{-20}$.\\
\indent Among the perfect entangling gates presented in Table \ref{tab:two-qubit_gates}, we only find zeros of the objective function for the CNOT gate, regardless of the biasing mode. In fact, the search produces many CNOT solutions with different times and bias values. Clearly, this method does not find all possible solutions since it does not reproduce the special case of the iSWAP solutions we found analytically above (this is partially due to the uniqueness of the bias values and long time required by the system to generate the iSWAP gate). However, the best CNOT solutions it does generate are faster, the shortest times being around $\approx 140\mathrm{ns}$. The Weyl chamber trajectories of the solutions are approximately on the tetrahedron's plane delimited by the points $(0,0,0)$, $(\pi,0,0)$, $(\pi/2,\pi/2,\pi/2)$, resulting in similar projections of the trajectories on the planes $(\gamma_x-\gamma_y)$ and $(\gamma_x-\gamma_z)$. The projection on the $(\gamma_x-\gamma_y)$ plane for each of the three biasing modes are shown in Fig. \ref{fig:4a}. The square-pulse parameters that generate those trajectories are $\varepsilon_{1,Right}\approx\varepsilon_{2,Right}\approx 3.344\mathrm{meV}$, $t_{Right}\approx 137\mathrm{ns}$, $\varepsilon_{1,Center}\approx\varepsilon_{2,Center}\approx 3.343\mathrm{meV}$, $t_{Center}\approx 142\mathrm{ns}$, $\varepsilon_{1,Outer}\approx 3.342\mathrm{meV}$, $\varepsilon_{2,Outer}\approx 3.344\mathrm{meV}$, $t_{Outer}\approx 141\mathrm{ns}$. Alternatively, when considering non-square pulses (see Appendix \ref{Non-square pulse model}) the time required to generate a CNOT gate is about $20-30\mathrm{ns}$ longer in comparison to the previous solutions.\\
\indent Notice that the bias terms $\varepsilon_i$ that gave us the shortest CNOT solutions are all close to the maximum allowed bias $\varepsilon_{max}\approx 3.344\mathrm{meV}$. This is a consequence of the time region used in the numerical search. In order to focus on the fastest possible gates under the bias constraints, we choose the lower bound to the time region of the search to be the minimum time required by a purely nonlocal Hamiltonian $H_{nonlocal}=\alpha(\varepsilon_1,\varepsilon_2)\sigma_z\otimes\sigma_z$ to generate a locally equivalent CNOT gate. That time, under the considered bias constraint, is given by $t_{min}=\frac{\pi}{4\vert\alpha(\varepsilon_{max},\varepsilon_{max})\vert}\approx 136\mathrm{ns}$.  We allow some room for the effects of the strong local part of the Hamiltonian by setting the upper bound of our search region to be 40ns greater. If we included even longer times, we would find additional solutions with lower bias values and, hence, lower coupling, since the gate time and coupling strength are inversely proportional. However, the 
weakness of the coupling (which depends on the bias at each qubit) and, consequently, long gate-times make the operations markedly sensitive to fluctuations of the bias and, to a much lesser degree, to fluctuations of the magnetic field gradient across the DQDs. For example, for a purely magnetic field gradient fluctuation, neglecting bias fluctuations, equal to $10^{-2}$ of the unperturbed value the infidelity of the entangling gate is $\sim5\times 10^{-5}$, whereas for a purely bias fluctuation greater than $10^{-4}$ of its unperturbed value the infidelity tends to its lower limit of $0.5$. In any case, due to the non-Markovian nature of the noise \cite{Kestner2013}, the decoherence of the system can in principle be suppressed by pulse sequence techniques \cite{Khodjasteh2009a,Khodjasteh2012,Kestner2013,Wang2014}.\\
\indent The approximate form of the Weyl chamber trajectories corresponding to a general Hamiltonian with weak Ising coupling is discussed in Ref. \onlinecite{Zhang2005}, where it is pointed out that when the local terms for the two qubits are symmetric, the trajectory in the Weyl chamber is approximately a straight line. This is indeed the case for the simpler Hamiltonian $H_0$ \eqref{Hamiltonian_z_vanished}, whose Weyl chamber trajectory is a straight line as shown in Fig. \ref{fig:3b}. On the other hand, when the coupling is weak and the local terms are not symmetric, the Weyl chamber trajectory of any single two-qubit operation is approximated by a sinusoidal curve that lies on the plane defined by $(0,0,0)$, $(\pi,0,0)$, $(\pi/2,\pi/2,\pi/2)$. This curve moves close to the line $(0,0,0)-(\pi,0,0)$ as seen in Fig. \ref{fig:4b}, making the CNOT gate $(\pi/2,0,0)$ a natural target. In fact, allowing long enough times, any combination of bias values in the weak coupling regime would eventually create an operation whose trajectory in the Weyl chamber would get close, yet not exactly equal, to the CNOT gate. However, the only way to exactly generate a CNOT gate is by a numerical search over bias values, as we have performed above.

\section{Conclusions}
Using the nonlocal properties of the capacitive-based two-qubit Hamiltonian \eqref{Two-qubit_Hamiltonian} and three different possible biasing modes, we have shown the type of entangling gates that can be directly generated by the two-qubit system. In particular, when tuning the bias of both DQDs in such a way that the two pairs of electrons at each DQD are slightly displaced away from their common center, there is a point where the local $\sigma_z\otimes I$ and $I\otimes\sigma_z$ terms of the Hamiltonian \eqref{Two-qubit_Hamiltonian} vanish. At this point, the Hamiltonian can directly generate a wide range of different entangling gates, including the iSWAP (or the locally equivalent DCNOT) gate, but not the CNOT (equivalently, CPHASE) gate. However, considering the complete two-qubit Hamiltonian \eqref{Two-qubit_Hamiltonian} and using a numerical approach we found that, in any of the three modes of biasing, the CNOT can be directly generated in a relatively short time. On the theoretical side, these results provide a useful characterization of the single-segment evolution of an important form of Hamiltonian. On the experimental side, these results are immediately practical, with relevance to ongoing experiments on capacitively coupled singlet-triplet qubits, and provide quantitative insight into the gate voltages that should be applied in order to generate maximal entanglement in minimal time in the presence of strong magnetic field gradients.\\

\section*{Acknowledgments}
The authors acknowledge support from the UMBC Office of Research through a Special Research Assistantship/Initiative Support award.

\begin{appendix}
\section{Single-qubit Hamiltonian}\label{app:Single-qubit Hamiltonian}
We use the Hund-Mulliken approximation for the orbital Hamiltonian of the biased qubit \cite{Burkard1999,Hu2000}, where only single-dot ground states (\textit{s} orbitals) are considered. In our approach the external magnetic field $\mathbf{B}$ is assumed along the $z$-axis and parallel to the plane ($x$-$z$) of the 2DEG, in contrast to Ref. \onlinecite{Burkard1999}, leading to a compression of the wave function perpendicular to the quantum dots' plane. Since the confinement is much stronger in the direction perpendicular to the 2DEG, which allows us to approximate the Coulomb interactions within a qubit using two-dimensional (2D) ($x$-$z$) integrals, we neglect the effect of the external magnetic field on the wave function.\\
\indent The two local minima of the quartic potential \eqref{Quartic_potential}, at ($\pm a, 0 $), are treated as isolated harmonic wells with ground-state wave functions:
\begin{equation}\label{Harmonic_ground-state_wavefunction}
\phi_{\pm a}(\mathbf{r})=\frac{1}{\sqrt{\pi}a_B}\exp\left\lbrace -\frac{1}{2a_B^2}\left[(x\mp a)^2+z^2\right]\right\rbrace,
\end{equation}
where $a_B=\sqrt{\hbar/m\omega_0}$. The orthonormalized single-particle states are given in Eq. \eqref{orthonormalized_single_particle_state}.\\
\indent The two-electron orbital Hamiltonian is composed of a single-particle Hamiltonian for each electron, the confinement potential, and the Coulomb interaction between electrons. The matrix elements of the orbital Hamiltonian are found by adding and subtracting the harmonic well potentials centered at ($\pm a,0$) \cite{Burkard1999}, in such a way
\begin{equation}
H_{orb}=h_{-a}^0(\mathbf{r}_1)+h_{+a}^0(\mathbf{r}_2)+W_{-}(\mathbf{r}_1)+W_{+}(\mathbf{r}_2)+C(\mathbf{r}_1,\mathbf{r}_2),
\end{equation}
where
\begin{align}
h_{\pm a}^0(\mathbf{r}_i)&=\frac{1}{2m}\mathbf{p}_i^2+\frac{m\omega_0^2}{2}\left[(x_i\mp a)^2+z_i^2\right],\\
W_{\pm}(\mathbf{r}_i)&=V(x,z)-\frac{m\omega_0^2}{2}\left[(x_i\mp a)^2+z_i^2\right],\\
C(\mathbf{r}_1,\mathbf{r}_2)&=\frac{1}{4\pi\kappa}\frac{e^2}{\vert\mathbf{r}_1-\mathbf{r}_2\vert}.
\end{align}
Here, $h_{\pm a}^0(\mathbf{r}_i)$ is the single-particle Hamiltonian plus the harmonic potential, $V(x,z)$ is the confinement potential given in Eq. \eqref{Quartic_potential}, and $C(\mathbf{r}_1,\mathbf{r}_2)$ is the Coulomb interaction between electrons with $\kappa$ being the dielectric constant of the host material (for GaAs $\kappa=13.1\epsilon_0$).\\
\indent The bias $\varepsilon$ is modeled as the energy difference between single-dot ground orbitals in the DQD created by an additional bias potential $V_b$ \cite{Petta2005}\cite{Stepanenko2007}. Its expression will depend on the ``tilting modality'' of the confinement potential, in such a way that the energy difference is always positive. In other words, if the DQD's potential is chosen to be tilted to the right, then the bias would be expressed as $\varepsilon=\langle \psi_{-a}\vert V_b\vert\psi_{-a}\rangle-\langle \psi_{a}\vert V_b\vert\psi_{a}\rangle$, whereas if the potential is tilted to the left the bias would be $\varepsilon=\langle \psi_{a}\vert V_b\vert\psi_{a}\rangle-\langle \psi_{-a}\vert V_b\vert\psi_{-a}\rangle$. With this in mind, the orbital Hamiltonian in the basis $\left\lbrace \vert S(2,0)\rangle,\vert S(0,2)\rangle,\vert S(1,1)\rangle,\vert T_0\rangle\right\rbrace$ is
\begin{equation}\label{Orbital_Hamiltonian_matrix}
H_{orb}=2\epsilon^0 +
\begin{pmatrix}
U+\varepsilon & X & -\sqrt{2}t^0 & 0\\
X & U-\varepsilon & -\sqrt{2}t^0 & 0\\
-\sqrt{2}t^0 & -\sqrt{2}t^0 & V_{+} & 0\\
0 & 0 & 0 & V_{-},
\end{pmatrix}
\end{equation}
where\cite{Burkard1999}
\begin{align}
\epsilon^0=&\langle\psi_{\pm a}\vert h_{\pm a}^0+W_{\pm a}\vert \psi_{\pm a}\rangle\\
=&\hbar\omega_0\left\lbrace 1+ \frac{3a_B^2}{32a^2}+\frac{3s^2}{8(1-s^2)}\left(1+\frac{a^2}{a_B^2}\right)\right\rbrace,\\
t^0=&\-\langle\psi_{\pm a}\vert h_{\pm a}^0+W_{\pm a}\vert \psi_{\mp a}\rangle-T\\
 =&\hbar\omega_0\left\lbrace\frac{3s}{8(1-s^2)}\left(1+\frac{a^2}{a_B^2}\right)\right\rbrace-T,\\
T=&\frac{1}{2}\langle\psi_a\psi_{-a}+\psi_{-a}\psi_a\vert C\vert\psi_{\pm a}\psi_{\pm a}\rangle \label{T},\\
U=&\langle\psi_{\pm a}\psi_{\pm a}\vert C\vert\psi_{\pm a}\psi_{\pm a}\rangle \label{U},\\
V_{+}=&\frac{1}{2}\langle\psi_a\psi_{-a}+\psi_{-a}\psi_a\vert C\vert\psi_a\psi_{-a}+\psi_{-a}\psi_a\rangle \label{Vplus},\\
V_{-}=&\frac{1}{2}\langle\psi_a\psi_{-a}-\psi_{-a}\psi_a\vert C\vert\psi_a\psi_{-a}-\psi_{-a}\psi_a\rangle \label{Vminus},\\
X=&\frac{1}{2}\langle\psi_{\pm a}\psi_{\pm a}\vert C\vert\psi_{\mp a}\psi_{\mp a}\rangle \label{X}.
\end{align}
Here $s=\exp[-a^2/a_B^2]$ is the overlap of the harmonic ground-state wave functions, $U$ is the on-site repulsion, $t^0$ is the extended hopping amplitude  between dots (which is composed by the single-particle tunneling amplitude and Coulomb interaction $T$), $V_{+}$ and $V_{-}$ are the Coulomb energies in the singly occupied singlet and unpolarized triplet states, respectively, and $X$ represents the coordinated hopping of both electrons between dots. The general closed-form expressions of Eqs. (\ref{T})-(\ref{X}) are given in Appendix \ref{app:Calculation_Coulomb_integrals}.\\
\indent Since $\varepsilon$ is chosen to be always positive, the Hamiltonian represented in Eq. \eqref{Orbital_Hamiltonian_matrix} corresponds to a confinement potential that is being tilted to the right, lowering the energy of the doubly occupied singlet $\vert S(0,2)\rangle$. An opposite scenario, i.e., the potential being tilted to the left instead, would be represented by the same matrix in Eq. \eqref{Orbital_Hamiltonian_matrix} but with the doubly occupied singlet states swapped in the matrix basis.\\
\indent At positive values of the bias $\varepsilon$, close to $\varepsilon_{max}$ (see Sec. \ref{sec:Two-qubit Hamiltonian}) where the coupling between qubits is stronger (see Fig. \ref{fig:2a}), the doubly occupied singlet state with energy $U+\varepsilon$ is far detuned from the other two singlets. In this light, we can neglect the far detuned singlet state and diagonalize the orbital Hamiltonian in the basis of the other two singlet states. Consequently, the lower eigenstate, when $\vert S(2,0)\rangle$ is the far detuned singlet state, reads as $\vert \tilde{S}\rangle =\sin\theta\vert S(0,2)\rangle+\cos\theta\vert S(1,1)\rangle$ (in contrast, if $\vert S(0,2)\rangle$ is far detuned from the other two singlet states, the lower eigenstate would be $\vert \tilde{S}\rangle =\sin\theta\vert S(2,0)\rangle+\cos\theta\vert S(1,1)\rangle$). In any case, the mixing angle $\theta$ introduced to parametrize the hybrid state $\vert \tilde{S}\rangle$ is defined as
\begin{equation}
\tan\theta=\frac{2\sqrt{2}t^0}{U-\varepsilon-V_{+}+\sqrt{\left(U-\varepsilon-V_{+}\right)^2+8\left(t^0\right)^2}},
\end{equation}
and the hybrid state's energy is given by
\begin{equation}
E_{-}(\varepsilon)=\frac{1}{2}\left[U-\varepsilon+V_{+}-\sqrt{\left(U-\varepsilon-V_{+}\right)^2+8\left(t^0\right)^2}\right].
\end{equation}
The exchange energy $J(\varepsilon)$ is the gap between the energies of the lowest singlet state, in this case the hybrid state, and the triplet state
\begin{equation}
J(\varepsilon)=V_{-}-E_{-}(\varepsilon).
\end{equation}
\section{Interaction Hamiltonian}\label{app:Interaction_Hamiltonian}
The elements of the interaction Hamiltonian \eqref{interactionHamiltonian} correspond to the Coulomb energy of the electrostatic interaction between electrons of different DQDs. Before we proceed, we must mention that the Coulomb interaction between DQDs creates additional terms to those included in the Hamiltonian \eqref{interactionHamiltonian}. The additional terms describe the coordinated hopping of the electrons in both DQDs, and supplementary corrections of the tunneling matrix element \cite{Stepanenko2007}. Nevertheless, these extra terms are at least five orders of magnitude smaller than the main energies and do not affect our results.\\
\indent The elements $V_{ij}$ in Eqs. (\ref{beta_1})-(\ref{alpha}) are defined in terms of three Coulomb integrals describing three possible electrostatic interactions between a pair of electrons from two identical DQDs. These integrals are \cite{Stepanenko2007}
\begin{equation}\label{2DQD_2electronInteraction}
\begin{aligned}
\mathcal{V}_n&=\langle\psi_{a,1}\psi_{-a,2}\vert C\vert\psi_{a,1}\psi_{-a,2}\rangle,\\
\mathcal{V}_m&=\langle\psi_{a,1}\psi_{a,2}\vert C\vert\psi_{a,1}\psi_{a,2}\rangle=\langle\psi_{-a,1}\psi_{-a,2}\vert C\vert\psi_{-a,1}\psi_{-a,2}\rangle,\\
\mathcal{V}_f&=\langle\psi_{-a,1}\psi_{a,2}\vert C\vert\psi_{-a,1}\psi_{a,2}\rangle,
\end{aligned}
\end{equation}
where $\mathcal{V}_n$ describes the interaction between an electron on the right QD of the left DQD and an electron on the left QD of the right DQD, $\mathcal{V}_m$ describes the interaction of an electron on the right (left) QD of the left DQD and an electron on the right (left) QD of the right DQD, $\mathcal{V}_f$ represents the interaction between an electron on the left QD of the left DQD and an electron on the right QD of the right DQD, and the subscript 1 (2) corresponds to the left (right) DQD.\\
\indent In the main text, we considered three possible biasing modes of the confinement potential of each DQD. Each of these modes have different representations of the Coulomb potentials $V_{ij}$. Starting with the mode where the confinement potentials in both DQDs are tilted to the same side (we choose the right side of each DQD), the Coulomb potentials are given by
\begin{equation}
\begin{aligned}
V_{\tilde{S}\tilde{S}}=&\sin^2\theta_1\sin^2\theta_2(4\mathcal{V}_m)\\
&+\sin^2\theta_1\cos^2\theta_2(2\mathcal{V}_n+2\mathcal{V}_m)\\
&+\cos^2\theta_1\sin^2\theta_2(2\mathcal{V}_m+2\mathcal{V}_f)\\
&+\cos^2\theta_1\cos^2\theta_2(\mathcal{V}_n+2\mathcal{V}_m+\mathcal{V}_f),\\
V_{\tilde{S}T_0}=&\sin^2\theta_1(2\mathcal{V}_n+2\mathcal{V}_m)\\
&+\cos^2\theta_1(\mathcal{V}_n+2\mathcal{V}_m+\mathcal{V}_f),\\
V_{T_0 \tilde{S}}=&\sin^2\theta_2(2\mathcal{V}_f+2\mathcal{V}_m)\\
&+\cos^2\theta_2(\mathcal{V}_n+2\mathcal{V}_m+\mathcal{V}_f),\\
V_{T_0T_0}=&\mathcal{V}_n+2\mathcal{V}_m+\mathcal{V}_f.
\end{aligned}
\end{equation}
When the confinement potential in each DQD is tilted towards the two-DQD center, the  Coulomb potentials have the form
\begin{equation}
\begin{aligned}
V_{\tilde{S}\tilde{S}}=&\sin^2\theta_1\sin^2\theta_2(4\mathcal{V}_n)\\
&+\sin^2\theta_1\cos^2\theta_2(2\mathcal{V}_n+2\mathcal{V}_m)\\
&+\cos^2\theta_1\sin^2\theta_2(2\mathcal{V}_n+2\mathcal{V}_m)\\
&+\cos^2\theta_1\cos^2\theta_2(\mathcal{V}_n+2\mathcal{V}_m+\mathcal{V}_f),\\
V_{\tilde{S}T_0}=&\sin^2\theta_1(2\mathcal{V}_m+2\mathcal{V}_n)\\
&+\cos^2\theta_1(\mathcal{V}_n+2\mathcal{V}_m+\mathcal{V}_f),\\
V_{T_0 \tilde{S}}=&\sin^2\theta_2(2\mathcal{V}_m+2\mathcal{V}_n)\\
&+\cos^2\theta_2(\mathcal{V}_n+2\mathcal{V}_m+\mathcal{V}_f),\\
V_{T_0T_0}=&\mathcal{V}_n+2\mathcal{V}_m+\mathcal{V}_f.
\end{aligned}
\end{equation}
Finally, the Coulomb potentials for a confinement potential in each DQD tilted away from the two-DQD center are
\begin{equation}
\begin{aligned}
V_{\tilde{S}\tilde{S}}=&\sin^2\theta_1\sin^2\theta_2(4\mathcal{V}_f)\\
&+\sin^2\theta_1\cos^2\theta_2(2\mathcal{V}_m+2\mathcal{V}_f)\\
&+\cos^2\theta_1\sin^2\theta_2(2\mathcal{V}_m+2\mathcal{V}_f)\\
&+\cos^2\theta_1\cos^2\theta_2(\mathcal{V}_n+2\mathcal{V}_m+\mathcal{V}_f),\\
V_{\tilde{S}T_0}=&\sin^2\theta_1(2\mathcal{V}_m+2\mathcal{V}_f)\\
&+\cos^2\theta_1(\mathcal{V}_n+2\mathcal{V}_m+\mathcal{V}_f),\\
V_{T_0 \tilde{S}}=&\sin^2\theta_2(2\mathcal{V}_m+2\mathcal{V}_f)\\
&+\cos^2\theta_2(\mathcal{V}_n+2\mathcal{V}_m+\mathcal{V}_f),\\
V_{T_0T_0}=&\mathcal{V}_n+2\mathcal{V}_m+\mathcal{V}_f.
\end{aligned}
\end{equation}

\section{Calculation of Coulomb integrals}\label{app:Calculation_Coulomb_integrals}
The integrals presented in Eqs.(\ref{T})-(\ref{X}) and \eqref{2DQD_2electronInteraction} have the general form $\langle\psi_{\alpha}\psi_{\gamma}\vert C\vert\psi_{\beta}\psi_{\delta}\rangle$. Here $\psi_j=N(\phi_{j_1}-g\phi_{j_2})$ is the orthonormalized single-particle state, Eq. \eqref{orthonormalized_single_particle_state}, where $N=\left(1-2sg+g^2\right)^{-1/2}$, $g=(1-\sqrt{1-s^2})/s$, $s=\langle\phi_{j_2}\vert\phi_{j_1}\rangle=\exp\left[-a^2/a_B^2\right]$, and $\phi_j$ is the harmonic potential ground-state wave function \eqref{Harmonic_ground-state_wavefunction}. Using $f_{ijkl}=\langle\phi_i\phi_k\vert C\vert\phi_j\phi_l\rangle$ to represent the two-electron Coulomb integral, we express the integral $\langle\psi_{\alpha}\psi_{\gamma}\vert C\vert\psi_{\beta}\psi_{\delta}\rangle$ in terms of the harmonic ground-state wave functions as
\begin{equation}
\begin{aligned}
\langle\psi_{\alpha}&\psi_{\gamma}\vert C\vert\psi_{\beta}\psi_{\delta}\rangle=N^4[f_{\alpha_1\beta_1\gamma_1\delta_1}\\
&-g\left(f_{\alpha_1\beta_1\gamma_1\delta_2}+f_{\alpha_1\beta_2\gamma_1\delta_1}+f_{\alpha_1\beta_1\gamma_2\delta_1}+f_{\alpha_2\beta_1\gamma_1\delta_1}\right)\\
&+g^2(f_{\alpha_1\beta_2\gamma_1\delta_2}+f_{\alpha_1\beta_1\gamma_2\delta_2}+f_{\alpha_1\beta_2\gamma_2\delta_1}\\
& \ \ \ \ \ \ +f_{\alpha_2\beta_1\gamma_1\delta_2}+f_{\alpha_2\beta_2\gamma_1\delta_1}+f_{\alpha_2\beta_1\gamma_2\delta_1})\\
&-g^3\left(f_{\alpha_1\beta_2\gamma_2\delta_2}+f_{\alpha_2\beta_2\gamma_1\delta_2}+f_{\alpha_2\beta_1\gamma_2\delta_2}+f_{\alpha_2\beta_2\gamma_2\delta_1}\right)\\
&+g^4f_{\alpha_2\beta_2\gamma_2\delta_2}].
\end{aligned}
\end{equation}

Since the harmonic ground-state wave functions $\phi$ (Eq. \eqref{Harmonic_ground-state_wavefunction}) are primitive 2D Gaussian functions, we use a slightly modified version of the method to evaluate integrals with \textit{1s} primitive Gaussians presented by Szabo and Ostlund in Ref. \onlinecite{Szabo1996} (we will not present here the details of the method and we refer the reader to Ref. \onlinecite{Szabo1996} for further information). The expression we found for the two-electron Coulomb integral is
\begin{equation}
\begin{aligned}
f_{ijkl}=&\langle\phi_i\phi_k\vert C\vert\phi_j\phi_l\rangle\\
 =&\frac{e^2}{4\pi\kappa}\sqrt{\frac{\pi}{2}}\frac{1}{a_B}\exp\left[-\frac{1}{4a_B^2}\left(\left(R_i-R_j\right)^2+\left(R_k-R_l\right)^2\right)\right]\\
 &\times\exp\left[-\frac{1}{16a_B^2}\left(R_i+R_j-R_k-R_l\right)^2\right]\\
 &\times I_0\left[\frac{1}{16a_B^2}\left(R_i+R_j-R_k-R_l\right)^2\right],
\end{aligned}
\end{equation}
where $I_0$ is the zero-order modified Bessel function of the first kind, $a_B$ is the effective Bohr radius, and $R_m$ is the distance from the two-DQD center to the center of the ground-state wave function $\phi_m$ (see Fig. \ref{fig:1}).

\section{Non-square pulse}\label{Non-square pulse model}
\begin{figure}[h]
    \centering
    \includegraphics[trim=0cm 4.7cm 0cm 5cm, clip=true,width=8.5cm, angle=0]{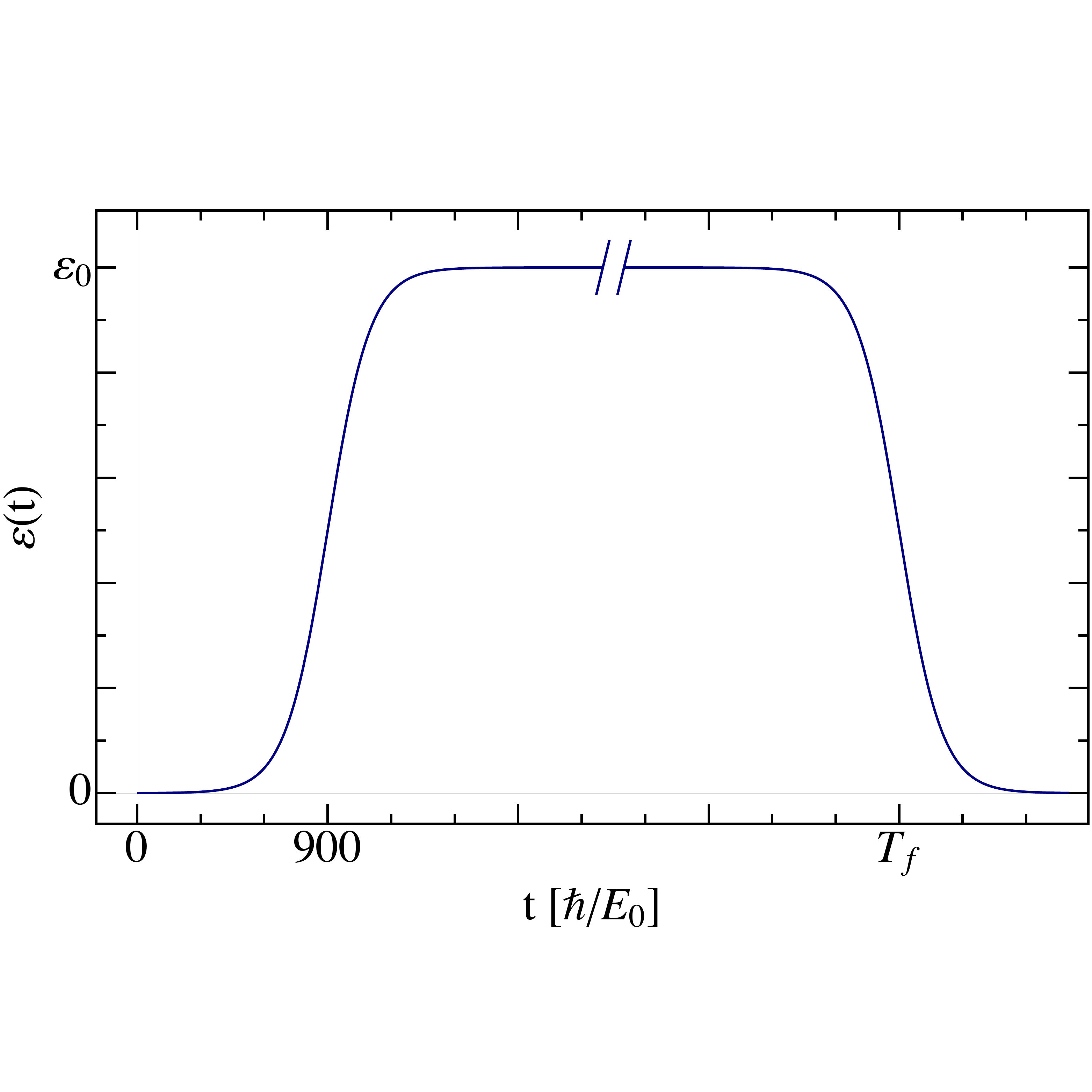}
    \caption{(Color online.) Curve shape of the time-dependent bias $\varepsilon(t)$, Eq. \eqref{eq:pulse_shape} (the time is in units of $\hbar/E_0$, $E_0$ is the confinement energy of the quantum dot).}\label{fig:5}
\end{figure}

To model a realistic rise and fall time of a non-square pulse, we use the following function for the bias $\varepsilon_i$ ($i=1,2$) in each qubit: 
\begin{equation}\label{eq:pulse_shape}
\varepsilon_i(t)=\varepsilon_{i,0}\left(\frac{1}{1+e^{\frac{-1}{100}(t-900)}+e^{\frac{1}{100}(t-T_f)}}\right),
\end{equation}
where $\varepsilon_{i,0}$ is the maximum value of the bias in the $i$th qubit (see Fig. \ref{fig:5}) and the constants were chosen in such a way that both rise and fall times are equal to $\sim 1.2\mathrm{ns}$ (currently the most common pulse generators' temporal resolution is of $0.8\mathrm{ns}$). The pulse starts at $t=0$ and ends at $t=T_f+900$ (this time is in units of $\hbar/E_0$, where $E_0$ is the confinement energy of the QD). Consequently, the Hamiltonian \eqref{Two-qubit_Hamiltonian} is now time-dependent and the system's evolution operator is the solution of Schr\"{o}dinger's differential equation $i\frac{\partial}{\partial t}U(t)=H(\varepsilon_1(t),\varepsilon_2(t))U(t)$. The numerical search for any realization of $U(t)$ that is locally equivalent to the desired entangling gate is pursued in the same fashion as in the time-independent case. The best-time solutions obtained are: $\varepsilon_{1,Right}\approx 3.339\mathrm{meV}$, $\varepsilon_{2,Right}\approx 3.344\mathrm{meV}$, $t_{Right}\approx 157\mathrm{ns}$, $\varepsilon_{1,Center}\approx3.336\mathrm{meV}$, $\varepsilon_{2,Center}\approx 3.344\mathrm{meV}$, $t_{Center}\approx 166\mathrm{ns}$, $\varepsilon_{1,Outer}\approx 3.344\mathrm{meV}$, $\varepsilon_{2,Outer}\approx 3.335\mathrm{meV}$, $t_{Outer}\approx 170\mathrm{ns}$.

\end{appendix}

\bibliographystyle{apsrev4-1}

\bibliography{library}

\end{document}